\documentclass[apj]{emulateapj}
 
\slugcomment{To appear in {\it{The Astrophysical Journal}}}
\shortauthors{Gonzalez et al.}
\shorttitle{Baryons in Clusters and Groups}

\newcommand \ho  {}
\newcommand{\kms}{~km~s$^{-1}$}

\newcommand\rfive{$r_{500}$}
\newcommand\rtwo{$r_{200}$}
\newcommand\mfive{{\fontfamily{pzc}\selectfont M}$_{500}$}

\begin{document}
\title{A Census of Baryons in Galaxy Clusters and Groups}
  
\author{Anthony H. Gonzalez\altaffilmark{1}, 
Dennis Zaritsky\altaffilmark{2,3}, \& Ann I.  Zabludoff\altaffilmark{2,3}}
\altaffiltext{1}{Department of Astronomy, University of Florida, Gainesville, FL 32611-2055}
\altaffiltext{2}{Steward Observatory, University of Arizona, 933 North Cherry
Avenue, Tucson, AZ 85721}
\altaffiltext{3}{Center for Cosmology and Particle Physics, Dept. of
Physics, NYU, New York, NY, 10003}
    
\begin{abstract}
We determine the contribution of stars in galaxies, intracluster stars, and
the intracluster medium to the total baryon budget in nearby galaxy clusters
and groups.  We find that the baryon mass fraction
($f_b\equiv\Omega_b/\Omega_m$) within \rfive\ is constant for systems with
\mfive\ between $6\times10^{13}\ \mathrm{M}_\odot$ and $1\times10^{15}\
\mathrm{M}_\odot$. Although $f_b$ is lower than the WMAP value, the shortfall
is on the order of both the observational systematic uncertainties and the
depletion of baryons within \rfive\ that is predicted by simulations.  The
data therefore provide no compelling evidence for undetected baryonic
components, particularly any that would be expected to vary in importance with
cluster mass.  A unique feature of the current analysis is direct inclusion of
the contribution of intracluster light (ICL) in the baryon budget.  With the
addition of the ICL to the stellar mass in galaxies, the increase in X-ray gas
mass fraction with increasing total mass is entirely accounted for by a
decrease in the total stellar mass fraction, supporting the argument that the
behavior of both the stellar and X-ray gas components is dominated by a
decrease in star formation efficiency in more massive environments.  Within
just the stellar component, the fraction of the total stellar luminosity in
the central, giant brightest cluster galaxy (BCG) and ICL (hereafter the
BCG+ICL component) decreases as velocity dispersion ($\sigma$) increases for
systems with $145 {\rm \ km\ s}^{-1} \le \sigma \le 1026$ km s$^{-1}$,
suggesting that the BCG+ICL component, and in particular the dominant ICL
component, grows less efficiently in higher mass environments.  The degree to
which this behavior arises from our sample selection, which favored systems
with central, giant elliptical galaxies, remains unclear.  A more robust
result is the identification of low mass groups with large BCG+ICL components,
demonstrating that the creation of ``intracluster" stars does not require a
massive cluster environment.  Within \rfive\ and \rtwo, the BCG+ICL
contributes on average 40\% and 33\% of the total stellar light, respectively,
for the clusters and groups in our sample. Because these fractions are
functions of both enclosed radius and system mass, care should be exercised
when comparing these values with other studies and simulations.
\end{abstract}

\keywords{galaxies: clusters: general --- galaxies:cD, formation, evolution, fundamental parameters}

\section{Introduction}
\label{sec:intro}

An important step in understanding any astrophysical system is a full
accounting of its constituent components.  Over scales ranging from groups to
clusters of galaxies, measuring how baryons are distributed among their
various gaseous and stellar phases can provide clues to the roles played by
AGN feedback, starburst winds, galaxy mergers, and tidal stripping
\citep{merritt1984,Roussel,valdarnini,delucia,kravtsov,cheng,ettori2006,monaco2006}
in driving galaxy evolution in these complex environments.  Determining the
total contribution of these baryonic components to the cluster mass is
similarly valuable.  While the baryon fraction in clusters may be expected to
reflect the universal value \citep[see][]{White1993,ettori2006}, baryon
accountings in clusters to date have fallen short of this expectation 
\citep{ettori2003,mccarthy} and uncovered possible
intriguing trends with cluster mass \citep{lin2003}.  With the precise
measurement of the universal baryon fraction from WMAP
\citep[$f_b\equiv\Omega_b/\Omega_m=0.176^{+0.008}_{-0.013}$;][]{spergel}, this
shortfall has gained in physical significance, leading to suggestions that
physical processes lower the cluster baryon fraction relative to the Universe
\citep[see, for example,][]{he}, that there may be significant, undetected
baryon components \citep[see][]{ettori2003,lin2004b}, or that WMAP
underestimates $\Omega_m$ \citep{mccarthy}. Any investigation along these
lines first requires measurement of {\it all} the baryon components, including
the hot, X-ray emitting gas and the stars in and out of galaxies.

One baryonic component that past studies were unable to include directly is
the intracluster stars (hereafter the intracluster light or ICL).  In recent
years, multiple studies have detected intracluster light via stacking of large
cluster samples \citep{zibetti2005}, deep observations of individual clusters
\citep{feldmeier2004,gonzalez2005,krick2006}, and even identification of {\it
individual} intracluster stars in very nearby clusters
\citep{durrell2002,feldmeier2004b,aguerri2005,ciardullo2005,gerhard2005}.  The
general consensus is now that all clusters have an ICL component.  The natural
extension of these studies is placing the ICL in the larger context of cluster
and group evolution, examining directly whether this component is important in
the chemical enrichment of the intragalactic medium
\citep{lin2004b,zaritsky2004}, the total stellar budget, or even in the total
baryon budget.  While recent work suggests that the ICL does not contribute
much to the overall baryon budget of the most massive clusters (the ones most
often observed), it could become important in lower mass systems where the
plasma is less dominant and the apparent discrepancy between observations and
the WMAP baryon fraction is most severe \citep[e.g.][]{lin2003}.  This issue
is mostly unexplored, as previous ICL measurements have come either from a few
systems with little dynamic range \citep{feldmeier2004b,krick2006} or from
stacked systems for which mass estimators have been crude \citep{zibetti2005}.

In our program we include the ICL's contribution to the total baryon mass
budget directly, explore the dependence of total baryon fraction on cluster
mass, and consider the relative importance of the stars in the ICL to those in
galaxies and of the total stellar mass to the total gas mass in the
intracluster medium (ICM).  We focus on a sample of 23 nearby galaxy clusters
and groups with ICL measurements from \citet{gonzalez2005} that span a wide
range in velocity dispersion, $145 {\rm \ km \ s}^{-1} \le \sigma \le 1026$
\kms.  We determine the cluster radii and masses at an overdensity of 500
times critical density (\rfive\ and \mfive;\S3.1), revisit our previous ICL
measurements (\S3.2), measure the total luminosity in the cluster galaxies
(\S3.3), determine stellar masses (\S3.4), and obtain gas masses from the
literature (\S3.5). We then derive relationships for the stellar and total
baryon mass fraction with cluster mass (\S4.1), examine the detailed behavior
of the ICL relative to the galaxy luminosity (\S4.2), and speculate about the
origin of the observed trends (\S4.3).  Throughout this paper we assume a
cosmology with $\Omega_M=0.27$, $\Omega_\Lambda=0.73$, and $H_0=70$ \kms.

\section{Sample Definition and Data}

In Paper I \citep{gonzalez2005}, we present a sample of 24 clusters and groups
for which we obtained drift scan imaging in Gunn $i$ using the Great Circle
Camera \citep{zaritsky1996} on the Las Campanas 1m Swope telescope. The sample
is comprised of systems at $0.03<z<0.13$ that contain a dominant brightest
cluster galaxy (BCG) with a major axis position that lies within 45$^\circ$ of
the direction of the drift scan, east-west, and span a range of velocity
dispersions and Bautz-Morgan types \citep{bautz1970}. We refer the reader to
Paper I for further details concerning the data and photometric reductions.

In Paper II \citep{Zaritsky2006}, we present velocity dispersions for 23 of
these clusters and groups.  Twelve of these $\sigma$'s are based on our own
observations with the Hydra spectrograph at the CTIO Blanco 4m telescope,
while the rest are derived from galaxy redshifts in the literature. For all
but four systems, the velocity dispersions are calculated using $>20$
confirmed members within 1.5 Mpc of the BCG. The system with the smallest
number of confirmed members has 8 galaxies and the lowest $\sigma$
($144^{+78}_{-57}$ \kms).  Details of the spectroscopic reductions and
velocity dispersion measurements are given in Paper II. The information from
both Papers I and II relevant to the current analysis is presented in Table
\ref{tab:data}.  The magnitudes in this Table do not include extinction or
k-corrections, which are listed in a separate column. When converting to
stellar masses in \S \ref{sec:stellarmass} we do include these corrections.

\begin{deluxetable*}{lcccccccc}
\tabletypesize{\scriptsize}
\tablecaption{Cluster Sample}
\tablewidth{0pt}
\tablehead{
\colhead{}       &\colhead{}  &\colhead{}     &\colhead{}        & \colhead{}           &\colhead{}  & \colhead{} &  \colhead{e+k}    & \colhead{BCG+ICL}   \\
\colhead{}       &\colhead{BM}  &\colhead{}     &\colhead{$\sigma$}        & \colhead{$\log$ \mfive}           &\colhead{M$_{BCG+ICL,r_{500}}$\tablenotemark{ab}}  & \colhead{M$_{gal,r_{500}}$\tablenotemark{c}} &  \colhead{Correction\tablenotemark{d}}    &               \colhead{Fraction}   \\
\colhead{Cluster}&\colhead{Type}&\colhead{$<z>$}&\colhead{(\kms)}&\colhead{(M$_\odot$)}&\colhead{(mag)} &\colhead{(mag)}     &\colhead{(mag)}   & \colhead{$<r_{500}$} 
}
\startdata
Abell 0122  & I    & .1134 & $677^{+108}_{-94}$        &   14.31  &$-25.60\pm0.04$&  $-26.11$ & 0.12 & $0.38\pm.04$  \\
Abell 1651  & I-II & .0847 & $990^{+112}_{-101}$       &   14.87  &$-25.63\pm0.10$&  $-26.88$ & 0.13 & $0.24\pm.04$ \\
Abell 2400  & II   & .0880 & $653^{+62}_{-57}\;\;$     &   14.26 &$-25.08\pm0.06$ &  $-26.10$ & 0.14 & $0.28\pm.06$  \\     
Abell 2401  & II   & .0571 & $459^{+101}_{-83}$        &   13.74 &$-24.61\pm0.03$ &  $-25.57$ & 0.09 & $0.29\pm.03$  \\     
Abell 2405\tablenotemark{e,f}& & .0368 & $145^{+78}_{-57}\;\;$   &   12.05  &$-23.68\pm0.14$ & $-21.24$ 	& 0.11 & $0.90\pm.14$   \\  
Abell 2571  & II   & .1091 & $ 671^{+83}_{-74}\;\;$&   14.30   &$-25.37\pm0.10$ &   $-25.53$  & 0.16 &$0.46\pm.10$ \\
Abell 2721  & II   & .1144 & $ 842^{+81}_{-74}\;\;$    &   14.63 &$-25.19\pm0.02$ &  $-26.90$ & 0.12 & $0.17\pm.02$ \\     
Abell 2730  & II   & .1201 & $1021^{+154}_{-135}\;\;\;$&   14.91 &$-26.04\pm0.11$ &  $-26.49$ & 0.11 & $0.40\pm.11$ \\     
Abell 2811  & I-II & .1079 & $860^{+114}_{-141}$       &   14.66 &$-25.64\pm0.08$ &  $-26.40$ & 0.11 & $0.33\pm.08$ \\     
Abell 2955  & II   & .0943 & $ 316^{+85}_{-67}\;\;$    &   13.19 &$-25.16\pm0.06$ &  $-24.96$ & 0.11 & $0.55\pm.06$ \\     
Abell 2969  & I    & .1259 & $982^{+192}_{161}$        &   14.85 &$-25.58\pm0.08$ &  $-26.62$ & 0.13 & $0.28\pm.08$ \\
Abell 2984  & I    & .1042 & $490^{+112}_{-91}$        &   13.84 &$-25.63\pm0.05$ &  $-25.51$ & 0.11 & $0.53\pm.05$ \\     
Abell 3112  & I    & .0753 & $941^{+139}_{-121}$       &   14.79 &$-25.76\pm0.03$ &  $-26.96$ & 0.08 & $0.25\pm.03$ \\     
Abell 3166  & I    & .1174 & $ 205^{+43}_{-36}\;\;$    &   12.56 &$-24.55\pm0.06$ &  $-23.99$ & 0.10 & $0.63\pm.06$ \\     
Abell 3693\tablenotemark{f}&&.1229&$1026^{+149}_{-130}\;$& 14.92 &$-25.49\pm0.05$ &  $-26.65$ & 0.12 &$0.26\pm.05$ \\   
Abell 3705  & III  & .0895 & $1009^{+81}_{-75}\;\;\;$  &   14.89 &$-24.37\pm0.02$ &  $-26.53$ & 0.15 & $0.12\pm.02$ \\     
Abell 3727  & III  & .1159 & $582^{+114}_{-95}$        &   14.09 &$-24.97\pm0.06$ &  $-25.83$ & 0.24 & $0.31\pm.06$ \\   
Abell 3809  & III  & .0623 & $544^{+55}_{-50}\;\;$     &   13.99 &$-24.69\pm0.04$ &  $-25.38$ & 0.08 & $0.35\pm.04$ \\     
Abell 3920  & I-II & .1268 & $ 426^{+136}_{-103}$      &   13.63 &$-25.06\pm0.04$ &  $-25.27$ & 0.12 & $0.45\pm.04$ \\     
Abell 4010  & I-II & .0955 & $ 634^{+145}_{-118}$      &   14.21 &$-25.56\pm0.08$ &  $-26.25$ & 0.11 & $0.35\pm.08$ \\     
APMC 020\tablenotemark{f}&&.1098&$ 242^{+57}_{-47}\;\;$&   12.80 &$-24.87\pm0.09$ &  $-23.55$ & 0.11 & $0.77\pm.09$ \\     
Abell S0084 & I    & .1080 & $ 522^{+155}_{-122}$      &   13.93 &$-25.38\pm0.04$ &  $-26.16$ & 0.12 & $0.33\pm.04$ \\ 
Abell S0296 & I    & .0696 & $ 477^{+76}_{-65}\;\;$    &   13.80 &$-25.19\pm0.04$ &  $-25.17$ & 0.09 & $0.51\pm.04$ 
\enddata
\tablenotetext{a}{For both the BCG+ICL and galaxies, the quoted absolute magnitudes do {\sl not} include the corrections for extinction or k-dimming
listed in column 8.}
\tablenotetext{c}{The estimated uncertainty for the galaxies is 0.15 mag.}
\tablenotetext{d}{ The extinction corrections are based upon the \citet{schlegel1998} dust maps and the k-corrections assume a passive
stellar population.}
\tablenotetext{e}{Abell 2405 is a superposition of two relatively poor systems at 11,000 and 27,000 km $s^{-1}$.  While this superposition hinders a robust measurement for the luminosity of the cluster galaxies, the BCG+ICL fraction is $>80\%$ within $r_{500}$ even if no galaxy background subtraction is performed.}
\tablenotetext{f}{No published Bautz-Morgan type. Abell 2405 and Abell 3693 each have two discrete redshift peaks.
Published Bautz-Morgan types do exist for these clusters, but not for the redshift peaks that we are studying.}
\tablecomments{The quoted uncertainties include the systematic uncertainties associated 
with the BCG+ICL luminosities due to sky subtraction, the background subtraction for
the galaxy luminosity, and the uncertainty in the faint end slope of the cluster galaxy
luminosity function. 
}
\label{tab:data}
\end{deluxetable*}

\section{Analysis}

To determine the ratio of baryons to total mass for each group and cluster in
our sample, we 1) estimate the total masses and cluster radii, 2) calculate
the combined luminosity of the brightest cluster galaxy and intracluster
light, 3) determine the total luminosity associated with the other cluster
members, 4) convert the luminosities of these components into stellar masses,
and 5) obtain X-ray gas mass fractions from the literature. In \S4, we discuss
the relationships between total system mass and stellar mass, X-ray gas mass,
and total baryon mass.

\subsection{Cluster Radii and Masses}

To estimate the total baryon fraction from the stellar components and ICM, we
must first choose a fixed radius within which the baryons can be summed and
then determine the baryonic and total masses within this radius. We choose
\rfive, the radius at which the cluster mass density exceeds the critical
value by a factor of 500, and the largest radius for which the current X-ray
data require no model extrapolation \citep{vikhlinin}.  Measurements of the
ICL typically do not extend even out to \rfive, but the light is sufficiently
concentrated within a few hundred kpc that the model extrapolation to \rfive\
introduces a minimal uncertainty.

The systems in our ICL sample have measured velocity dispersions, but generally
lack X-ray data. To obtain \rfive\ and \mfive\ values for the clusters in
our ICL sample that are directly
comparable to those determined in X-ray studies, we therefore derive
calibrations of the $\sigma$-\rfive\ and $\sigma$-\mfive\ relations using the
clusters from \citet{vikhlinin} that also have published velocity dispersions
\citep{girardi,wu}.  The fit to the $\sigma$-\rfive\ relation has a slope
consistent with the theoretical expectation ($1.07\pm 0.12$ vs 1) and
corresponds to a cluster with $\sigma=1000$ \kms\ having an \rfive\ of 1.41
Mpc.  The best fit to the $\sigma$-\mfive\ relationship using the
\citet{vikhlinin} sample has a slope consistent with the theoretical
expectation ($3.37 \pm 0.53$ vs. 3) and corresponds to a cluster with $\sigma
= 1000$ \kms\ having an enclosed mass, \mfive, of $7.6 \times 10^{14}
M_\odot$.  A similar analysis can be performed for \rtwo\ using the sample of
\citet{arnaud2005}, but with the limitation that only five clusters in this
sample have published velocity dispersions. Fixing the slope to the value
derived for the \citet{vikhlinin} sample, the resulting normalization yields
an \rtwo\ of 2.2 Mpc for a $\sigma=1000$ \kms. We utilize this \rtwo\ relation
in \S\ref{sec:vdrad} when quantifying the radial dependence of the ICL.

For comparison, we also derive \rfive\ and \rtwo\ using the approach of
\citet{hansen2005}, directly calculating the radius within which the number
density of cluster galaxies implies that the mass density exceeds the critical
value by a factor of 200 or 500. If the bias between the galaxy distribution
and dark matter is not large, this approach is expected to yield a reasonable
estimate for these radii.  As in \citet{hansen2005}, we use the
\citet{blanton2003} SDSS $i-$band luminosity function integrated over redshift
to compute the field density \citep[see][]{hansen2005}.  When computing \rtwo\
and \rfive, we impose a faint magnitude limit of $m_I=18$; varying this limit
by $\pm0.5$ mag alters the derived \rtwo\ values by $\pm5$\%. We find that two
approaches yield consistent values except for clusters with low velocity
dispersions ($\la 400$ \kms), where the \citet{hansen2005} approach yields
larger radii (by as much as a factor of two). For all analyses below we use
the X-ray derived values; no results in the paper change qualitatively if we
instead use the values from the \citet{hansen2005} approach.

\subsection{Total Luminosity of BCG+ICL}

A key question in previous discussions of these data
\citep{zaritsky2004,gonzalez2005} was whether the BCG and ICL are distinct
entities.  A central result of Paper I is that the stellar surface brightness,
position angle, and ellipticity profiles of these clusters are best reproduced
when the BCG and ICL are fit as separate components.  In this analysis we
choose to combine the luminosity of these two components and refer to the
combined entity as simply the BCG+ICL.  There are several strong motivations
for this choice in the current context.  Observationally, the separation of
the two components is difficult and the combined luminosity is therefore a
more robust quantity than the luminosity of either separately.  Using the
combined quantity also avoids choosing among the various ICL definitions in
the literature \citep[e.g.,][]{feldmeier2004,zibetti2005}, enabling a more
straightforward comparison between our results and other observational studies
and simulations.  Lastly, because we are interested here in a full baryon
accounting, the question of whether they reside in the central galaxy or in
intracluster space is not as critical as in other contexts.  In later
discussions of the ICL relative to the galaxy luminosities (\S
\ref{sec:stellar}), where the distinction between BCG and ICL is relevant, we
reiterate from Paper I that the bulk ($> 80$\%) of the luminosity in the
cluster's BCG+ICL is contained in the ICL.

We determine the luminosity of the BCG+ICL component as follows.  In Paper I
we used the GALFIT package \citep{peng2002} to fit the surface brightness
distribution in each cluster out to a radius of 300 \ho kpc from the BCG.  The
BCG and ICL were best fit with separate $r^{1/4}$ profiles.  Here we use those
best-fit profiles to construct a 2-D model image from which we determine the
flux within circular apertures corresponding to \rfive\ (in our analysis of
the baryon mass fraction; \S4.1) and to fractions of \rtwo\ (in our
determination of the BCG+ICL's contribution to the total stellar light;
\S4.2).  We estimate the systematic errors in the measured fluxes using the
$\pm1\sigma$ systematic uncertainties in the structural parameters of the
best-fit profiles (these exceed the statistical errors due to uncertainties in
the sky level; see Paper I).  The advantages of using the models rather than
placing apertures on the original data are that we minimize contamination from
cluster galaxies, interpolate directly over masked regions, and provide a
robust means of quantifying the systematic uncertainties.  We use circular
rather than elliptical apertures to enable direct comparison with other
studies and simulations.

An implicit assumption in this work is that our models from Paper I, which are
determined for $r<300$ kpc, adequately describe the ICL at all radii. In our
models, 80\% of the total light of the BCG+ICL is contained in the central 300
kpc, making it unlikely that we are significantly overestimating the
luminosity of this component with our extrapolation.  Moreover, the BCG+ICL
dominate the total cluster luminosity within this radius, and for our data the
potential bias from inclusion of faint, unresolved cluster galaxies in the ICL
is at most a few percent \citep{gonzalez2000}.  Although the uniformity of the
profiles within 300 kpc argues that extrapolating is reasonable, we will
underestimate the ICL luminosity if there is any significant contribution at
larger radii that is not included in our model. We expect that the best method
of testing this assumption is via extension of current planetary nebulae
studies in nearby clusters like Virgo \citep{feldmeier2004b} to large
fractions of \rtwo. Because of the nature of this study, we are most concerned
with overestimating the ICL, and large overestimates ($>$ 30 \%) are not
possible.

\subsection{Total Luminosity of Cluster Galaxies}
\label{sec:luminosity}
We compute the total luminosity of non-BCG cluster galaxies within the same
apertures (\rfive\ and fractions of \rtwo) used for the BCG+ICL.  Hereafter,
we refer to the non-BCG cluster galaxies as simply ``cluster galaxies".  After
excluding stars using the SExtractor \citep{Bertin1996} stellarity index, we
sum the flux of all galaxies fainter than the BCG and brighter than $m_I=18$
lying within the aperture. We then perform a statistical background
subtraction, and include a completeness correction to account for the
contribution of fainter cluster members (see below).  In cases where the
aperture size is larger than the width of the drift scan images, we also apply
a correction to account for the lost area. Specifically, we weight the
luminosity contribution of each galaxy by the inverse of the fractional area
lost at that radius (e.g., a galaxy would have a weighting factor of two if
half the area is lost).

Because the determination of the flux density from background galaxies is
critical in this type of measurement, we estimate the background using two
independent methods.  First, we compute the background galaxy density with an
annular region located $30\arcmin$ to $60\arcmin$ from the BCG.  This method,
which was also used in Paper I, has the advantage that the photometry for the
background and cluster galaxies is performed in the same fashion from the same
images.  However, large scale structure may bias the observed background level
even at these large radii. To ascertain whether we are suffering from such a
bias, we also compute the background level using the Sloan Digital Sky Survey
luminosity function from \cite{blanton2003}, integrating over redshift and
applying mean evolutionary and k-corrections derived from \citet{bruzual2003}
models with the Padova isochrones \citep{padova1994}.  A comparison of the
results from the two techniques shows that for the ensemble of systems the
total cluster galaxy luminosities after background subtraction are consistent
to within 10\%. In the subsequent discussion, we apply the first method of
direct background subtraction because this approach relies on fewer
assumptions.

After background subtraction, we include a completeness correction to account
for the luminosity contribution from cluster members fainter than the limiting
magnitude. For this correction, we adopt the cluster luminosity function of
\citet{christlein2003}, $\alpha=-1.21$ and $M_{*,R}=-21.14$, using $R-I=0.82$
to convert from $R$ to $I$. Although the value of $M_*$ is identical to the
$i-$band value for the field from \cite{blanton2003}, the cluster luminosity
function has a slightly steeper faint end slope.  Using a different value of
$\alpha$ systematically shifts the total luminosity due to cluster galaxies
and thus the total stellar light of the cluster.  For $\alpha=-1$, the
computed total luminosity of the cluster galaxies would on average be 12\%
lower, increasing the relative importance of the BCG+ICL, but decreasing the
stellar contribution to the overall baryon budget.  The total luminosities of
the BCG+ICL and of the cluster galaxy components are presented in Table
\ref{tab:data}.

\begin{figure*}
\plotone{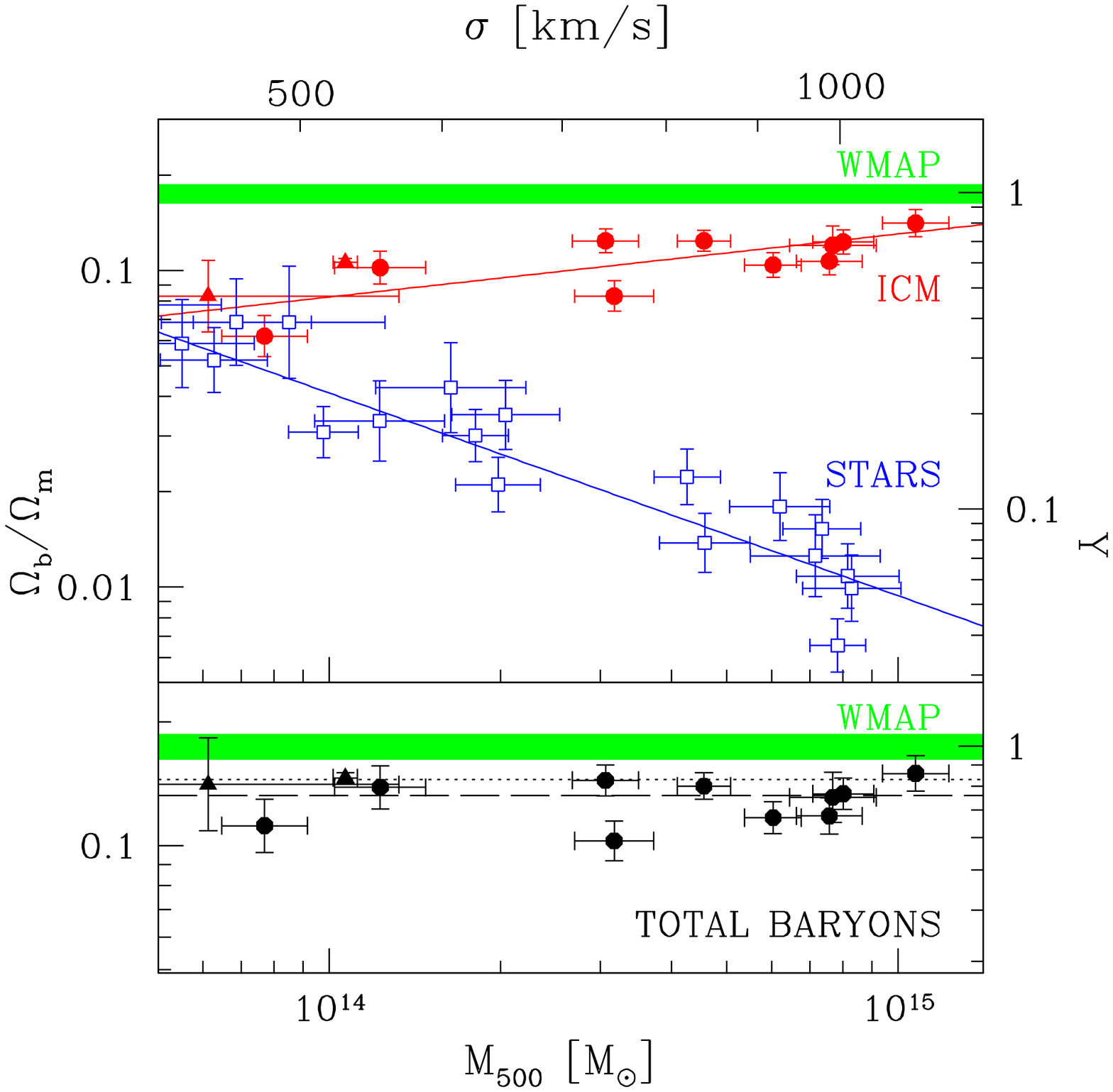}
\caption{Determination of cluster and group baryon fractions within \rfive\ as
a function of \mfive\ (bottom axis) and velocity dispersion (top axis).  
Plotted are X-ray gas mass fractions from \citet{vikhlinin} (filled circles) and 
\citet{gastaldello2006} (filled triangles), and the stellar mass fractions
(BCG+ICL+galaxies, open squares) for systems in our sample with masses that overlap
the range shown for the X-ray studies. All measurements are within \rfive.
Overplotted are the best fit relations
for the \citet{vikhlinin} sample gas mass fractions and for our stellar mass
fractions. The WMAP 1$\sigma$ confidence region for the universal baryon
fraction from \citet{spergel} is shown for comparison, and the right-hand axis
shows $Y$, the ratio of the baryon fraction for each component to the
universal value from WMAP.  
{\it Lower panel:} The total baryon fraction derived for the \citet{vikhlinin}
and \citet{gastaldello2006} clusters if our best-fit stellar baryon relation
is used to estimate a stellar baryon contribution for each of these
systems. The weighted mean for the sample (dashed line) is $\Omega_b/\Omega_m=0.133\pm0.004$.  
We observe no trend in baryon fraction with cluster mass. The error bars on the individual 
data points and the weighted mean include only statistical uncertainties. 
The unweighted mean for the combined \citet{zhang2007} and \citet{rasmussen2006}
samples (dotted line) is included to provide a sense of the systematic uncertainties.
\label{fig:baryons}
}
\epsscale{1.0}
\end{figure*}

\subsection{Total Stellar Masses}
\label{sec:stellarmass}
The conversion from luminosity to stellar mass for the cluster galaxies and
the BCG+ICL potentially incurs the most significant uncertainty in the total
baryon mass budget.  A straightforward theoretical calculation, based on
PEGASE2.0 \citep{pegase} for a Salpeter IMF and a 10 Gyr old population,
yields $M/L_I = 4.7$, but recent observations indicate that real systems
exhibit lower $M/L$.  To estimate a mean $M/L$, we use the SAURON results from
\citet{Cappellari2006}, which provide an empirical determination of the
central stellar $M/L$ using Schwarschild dynamical modeling of two-dimensional
kinematic data.  Equation 9 in \citet{Cappellari2006} provides the luminosity
dependence of the mass-to-light ratio in the $I-$band.  We compute a
luminosity-weighted M/L for $L>0.25 L_*$ (the range over which the SAURON
relation is determined) using the same \citet{christlein2003} luminosity
function as in \S\ref{sec:luminosity}. We adopt the resultant value,
$M/L=3.6$, for both the cluster galaxies and the BCG+ICL in our baryon
calculation, but note that the uncertainty is at least 10\% and that $M/L$
varies with both mass \citep{Cappellari2006,Zaritsky2006} and stellar
population.  An improved accounting for the ICL component would require
knowledge of the source of the ICL stars if the variations in $M/L$ are driven
by population, rather than dynamical, differences. An improved accounting for
the cluster galaxy population would require determining the fraction of spiral
galaxies in these clusters and using a more appropriate $M/L$ for the spirals.

\subsection{Total Mass in the Intracluster Medium}
Determination of total cluster baryon fractions also requires measurement of
the mass of hot gas in the intracluster medium. Because we lack these data for
clusters in our sample, in \S \ref{sec:results} we fit the behavior of the
stellar mass fraction with cluster mass and use this relation to derive total
baryon fractions for clusters with published X-ray gas fractions.  For the gas
mass fractions we adopt the values of X-ray gas fraction inside of \rfive ,
$f_{g}$, and of enclosed mass, \mfive, from the tabulations by
\citet{vikhlinin} and \citet{gastaldello2006}. The latter adds only two new
systems, but helps constrain the baryon fraction at low mass.  We note that
there are several systems for which the X-ray gas fraction was measured by
both groups.  In these instances there is reasonably good agreement, and hence
we expect no significant relative systematic biases between the two samples.

\section{Results and Discussion}
\label{sec:results}

\subsection{Baryon Mass Fraction versus Cluster Mass}
\label{sec:baryonfraction}

In this section, we obtain two principal results.  First, summing the three
baryonic components --- the hot gas of the intracluster medium, the stars of
the BCG+ICL, and the stars of the other cluster galaxies --- gives a total
baryon mass fraction, $f_b$, that is constant as a function of total mass
(Figure 1). This constancy contradicts previous studies and arises in large
part from the inclusion of the ICL for systems with \mfive\ $<$ few $\times
10^{14} M_\odot$.  Second, the weighted mean value of the total baryon mass
fraction is $f_b=0.133$.\footnote{Defining $Y$ to be the ratio of the baryon
fraction to the cosmic value from the WMAP third year results, this value for
the total baryon fraction corresponds to $Y=0.76$.}  For reasons discussed
below, we conclude that there is no compelling case for undetected baryons in
the groups and clusters in our sample.

We fit the behavior of the stellar and X-ray gas mass fractions with mass as
power laws (top panel of Figure \ref{fig:baryons}).  Within \rfive, the stars
result in a relationship $\log(f_{*}) = (7.57 \pm 0.08) -
(0.64\pm0.13)\log$(\mfive) and the plasma in $\log(f_{g}) = (-3.87 \pm
0.04) + (0.20 \pm 0.05)\log$(\mfive) for the sample of \citet{vikhlinin}.\footnote{Note
that while this fit applies only to the \citet{vikhlinin} data set, the \citet{gastaldello2006} points
in Figure \ref{fig:baryons} are consistent with this relation.}
fit the \citet{vikhlinin} sample alone  The behavior of the two components
is clearly inverted. The plasma component, which dominates at cluster masses
$> 10^{14} M_\odot$, has a shallower dependence on mass than the stellar
component.  We use our best fit relation for the stellar mass to convert the
X-ray gas fractions from \citet{vikhlinin} and \citet{gastaldello2006} to
total baryon fractions on a cluster-by-cluster basis. The resulting total
baryon fractions, shown in the lower panel of Figure \ref{fig:baryons}, are
independent of cluster mass over the range of \mfive\ from $6\times10^{13}$ to
$10^{15}$ M$_\odot$. If one considers only the statistical uncertainties
associated with our total cluster baryon fraction and the WMAP measurement,
then the two values are discrepant at the level of 3.2$\sigma$.

There are three possible interpretations of this result in the context of a
full baryon accounting. First, systematic uncertainties may be significantly
larger than the random errors, and our measurement may therefore be consistent
with the universal WMAP value. Because the gas mass fraction is much larger
than the stellar mass fraction over the majority of the mass range we probe,
we focus this discussion on potential errors in the gas mass
fraction. Systematic errors, by their nature, are often difficult to calculate
and can best be illuminated by comparing independent studies.
\citet{zhang2007} and \citet{rasmussen2006} provide gas mass fractions for
high and low mass systems, respectively.  \citet{zhang2007} find gas fractions
that are 15\% larger than those we adopted over the range probed by their
sample. \citet{rasmussen2006} also find gas fractions that are larger than
those we adopted for comparably low mass systems.  Together, these two samples
cover the mass range of our adopted sample and result in an average
$f_b=0.149$, which is $2.1\sigma$ discrepant with the WMAP value and 12\%
larger than the $f_b$ we derive for the \citet{vikhlinin} data set.  We
conclude that indeed the systematic uncertainties dominate, that the sense of
the uncertainty works to diminish the discrepancy between measurements and
expectations, and that the case for undetected baryon components rests on
removing these systematic uncertainties.  Second, the baryon mass fraction
with \rfive\ in clusters may be different from the universal value.  Cluster
simulations routinely predict baryon fractions that are lower than the
universal baryon fraction by $\sim$10\%.  We have deliberately not adopted any
of those as the target because there are still a minority of simulations that
predict a baryon fraction larger than universal
\citep[e.g.][]{kravtsov}. Third, the shortfall may be resolved by yet
undetected baryon components. The constancy of the measured baryon fraction
with total system mass argues against a large undetected component, but one
could contribute at a modest level and bypass current detection.  In the end,
all of these possibilities may contribute at some level to a final resolution
of the baryon accounting, but at this point we are left to conclude that there
is no compelling evidence for undetected baryons.

We close this section by comparing our study to that of \citet{lin2003}, who
obtained a dependence of the baryon fraction on cluster mass in conflict with
our result.  The range of masses covered by the two studies is the same, and
we both use \rfive\ as our fiducial radius, so the discrepancy is real.  The
most obvious differences are our inclusion of the ICL component and our
different normalization of the gaseous component.  We model the effect of
including the ICL into their accounting by using our relationship for the
BCG+ICL mass vs. system mass and ``correcting'' their values of the baryon
fraction on a cluster-by-cluster basis.\footnote{This is a slight ($\sim$20\%)
overcorrection because it includes the BCG, which \cite{lin2003} had already
included.} The relationship between baryon fraction and mass decreases from a
$\sim$3$\sigma$ result in the original \citet{lin2003} study to a 1.8$\sigma$
result in the corrected case. Therefore, the inclusion of the ICL explains
part of the discrepancy between our result and that of \citet{lin2003},
although apparently not all of it. The lack of the ICL in the \citet{lin2003}
accounting was not an oversight --- they attempted to model the effect of the
ICL in \cite{lin2004b}, but simply lacked data of the sort we present here.
While the exclusion of the ICL causes a relative underprediction of the
contribution of stellar baryons for the lowest mass sytems, the Lin et
al. adopted X-ray gas fractions, which are $\sim 25$\% larger than ours and
hence consistent with the WMAP baryon mass fraction for the most massive
systems, cause a relative overprediction of baryons in the highest mass
systems.  These two effects lead to the apparent mass-dependent baryon
fraction of Lin et al.  As with the comparison among different studies
discussed previously, we have an unresolved discrepancy in the gas mass
fractions at the $\sim 20$\% level that highlights the difficulty in such an
analysis and supports our contention that there is yet no compelling evidence
for either additional baryonic components or a trend in baryon fraction with
system mass.

\begin{figure}
\plotone{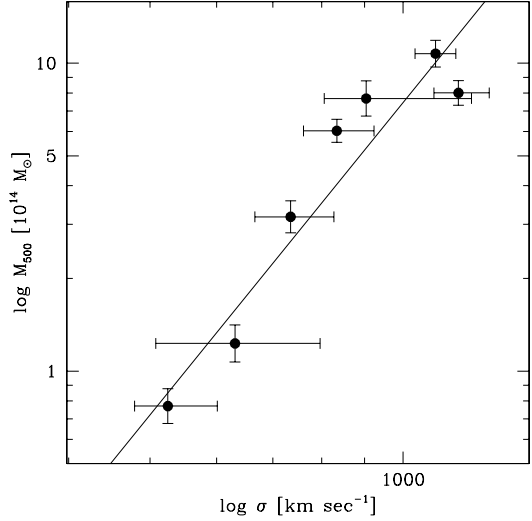}
\caption{Calibration of velocity dispersion to X-ray derived masses within
\rfive. We plot the seven systems from \cite{vikhlinin} for which we found
velocity dispersions in the literature \citep{girardi,wu}. The line is our fit
to the data, and hence our calibration for converting the velocity dispersions
measured for our optical clusters into \mfive 's.  }
\label{fig:masscal}
\end{figure}

\subsection{Stellar Component of Clusters and Groups}
\label{sec:stellar}
\subsubsection{Behavior vs. \mfive}

We now consider the behavior of the stellar component in more detail.  We
compute the dependence of the optical luminosity on \mfive\ for the cluster
galaxies, the BCG+ICL, and the sum of these populations.  There are multiple
studies of the dependence of the total stellar luminosity of cluster galaxies
on cluster mass.  In the $K-$band, \citet{lin2004a} find that for the galaxies
alone (excluding the BCG) $L\propto M^\gamma$ with $\gamma=0.82\pm0.05$
($\gamma=0.82\pm0.04$) within \rfive\ (\rtwo), and \citet{ramella2004} find
$\gamma=0.74\pm0.06$ within \rtwo.  A least-squares fit to our $I-$band data
for the cluster galaxies yields $\gamma=0.71\pm0.07$ within \rfive, which is
consistent with these results and most other published values \citep[][but see
\citet{kochanek2003} for an exception]{girardi2000,rines2004}.  We conclude
that the conflict between our result --- that the baryon mass fraction is
insensitive to system mass for groups and clusters --- and those of past
studies does not arise predominantly from our measurement of cluster galaxy
properties.

\begin{figure}
\plotone{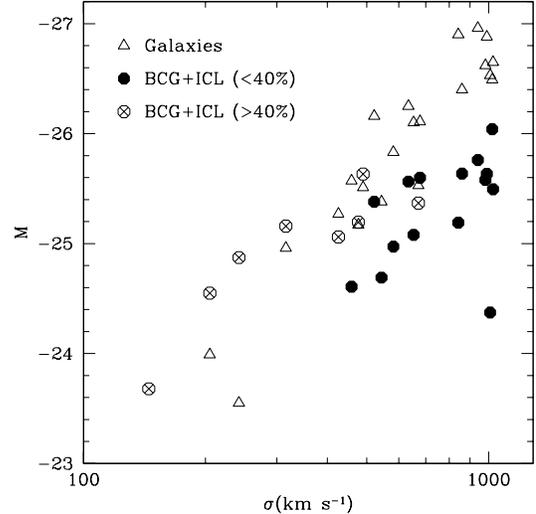}
\caption{The total absolute magnitude of the BCG+ICL component (circles) and
of cluster galaxies (triangles) within $r_{500}$ as a function of cluster
velocity dispersion. The filled circles denote systems where the BCG+ICL
contributes less than 40\% of the total luminosity within this radius, while
open-crossed circles correspond to systems with higher fractions of light in
the BCG+ICL.  The BCG+ICL luminosity depends less strongly on the velocity
dispersion than does the total galaxy luminosity. We also note that there is
no systematic difference in the BCG+ICL absolute magnitudes for clusters with
high and low BCG+ICL fractions. The BCG+ICL fraction differences are instead
driven predominantly by variations in the total absolute magnitudes of the
non-BCG cluster galaxy populations.
\label{fig:magsig}}
\end{figure}                

\begin{figure}
\plotone{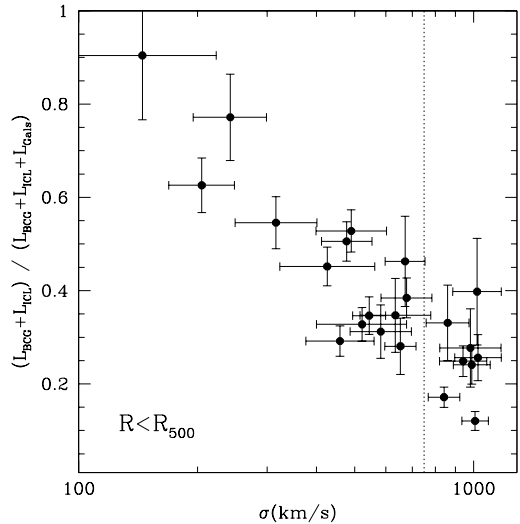}
\caption{The fraction of the total stellar luminosity, measured within \rfive,
which is contained in the BCG+ICL component as a function of cluster velocity
dispersion.  The dotted vertical line corresponds to $\sigma=750$ \kms,
roughly the median velocity dispersion for an Abell richness class $R\ge1$
cluster \citep{zabludoff1990}.
\label{fig:icl}}
\end{figure}

In comparison with the cluster galaxy luminosity versus cluster mass relation,
the correlation between BCG+ICL luminosity and cluster mass is weaker,
increasing by only about a factor of two in total luminosity between poor
groups with $\sigma=400$\kms~and rich clusters with $\sigma=1000$\kms~(Figure
\ref{fig:magsig}).  The existence of groups in which the fraction of stars in
the BCG+ICL exceeds that in galaxies out to \rfive\ (Figure \ref{fig:icl})
demonstrates that the rich cluster environment is not a requirement for the
production of a significant intracluster stellar population.  Instead, it
appears that local density maxima in both the group and cluster environments
are capable of generating intracluster stars, presumably via tidal processes.
This conclusion is valid whether one is interested in the combination of the
BCG and ICL, or in the ICL alone, because the latter contains the bulk
(typically $>80$\%) of the total light in the two components
\citep{gonzalez2005}. However, our results should not be interpreted to mean
that all groups have a significant ICL component, because our original sample
selection focused on systems with dominant BCGs.

The observed weak dependence in the BCG+ICL is given by
$\gamma=0.39\pm0.05$. Physically, this result implies that production of the
BCG+ICL --- presumably from the stripping of stars from galaxies --- is more
efficient in group (or subcluster) environments than in clusters.  Again, we
caution that when selecting clusters and groups for this sample, we
specifically targeted systems with dominant BCGs. If there is a strong
correlation between the presence of a dominant BCG (i.e., our sample) and the
total BCG+ICL luminosity (which is not necessarily the case because the ICL
contributes much more light), then our selection criteria would artificially
produce a smaller $\gamma$ than that actually found in nature.

\subsubsection{Total Cluster Luminosities and Mass-to-Light Ratios}

The observed weak dependence of the BCG+ICL luminosity on cluster mass, if
real, implies that $\gamma$ for the total stellar light within \rfive\
($L_{500}$) must be smaller than the value for the galaxy population alone.
The best fit value for the behavior of $L_{500}$ is $\gamma=0.47\pm0.05$, with
a scatter of only 0.26 mag about the best fit. Recasting in terms of a
mass-to-light ratio yields $M_{500}/L_{500} \propto L_{500}^{1.13\pm0.23}$.
For comparison, literature values that include only the cluster galaxy
contribution typically have $M/L\propto L^{0.3}$.  Thus, if the slope we
measure for the BCG+ICL component is not significantly biased by our exclusion
of groups without a dominant BCG, then our data argue that galaxy cluster
mass-to-light ratios are much more strongly dependent upon system mass than is
commonly presumed.

Conversely, if our selection against groups lacking a dominant BCG is the
reason that the BCG+ICL and cluster galaxy luminosities have different
observed cluster mass dependences, then there must exist a correlation between
mass-to-light ratio and the presence of a dominant BCG. Specifically, for a
given velocity dispersion, groups with a dominant BCG have higher BCG+ICL
luminosities and hence lower mass-to-light ratios.

\subsubsection{BCG+ICL Luminosity Fraction:  Dependence on Velocity Dispersion and Radius}
\label{sec:vdrad}

We determine the fraction of total stellar luminosity contained in the BCG+ICL
component as a function of velocity dispersion and radius.  In Figure
\ref{fig:icl}, we plot the BCG+ICL luminosity fraction, $L_{BCG+ICL}/
(L_{gal}+L_{BCG+ICL})$ as a function of velocity dispersion within \rfive,
which is slightly larger than the physical region within which the BCG and ICL
were modeled in Paper I. We find a trend of decreasing BCG+ICL luminosity
fraction with increasing velocity dispersion. The same trend is observed if we
use a fixed physical radius of 300 kpc, but the scatter is higher.  The
presence of this trend is also independent of whether \rfive\ or a fixed
fraction of \rtwo\ is used for the abscissa.

\begin{figure}
\plotone{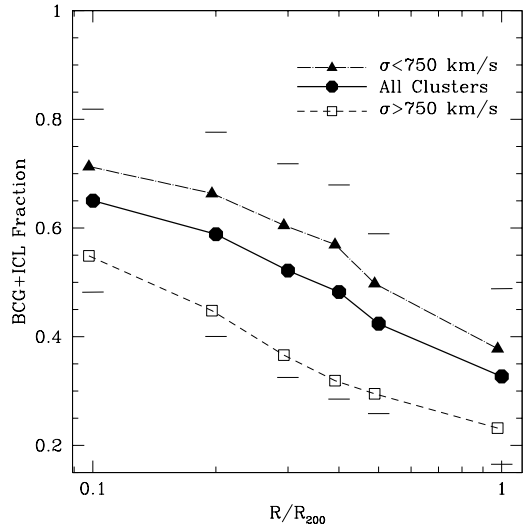}
\caption{The cumulative fraction of light contained in the BCG+ICL as a
function of radius. Subdividing the sample into clusters with $\sigma>750$
\kms~(open boxes) and $\sigma<750$\kms~(filled triangles) demonstrates that
the BCG+ICL fraction is systematically lower in the more massive clusters. The
horizontal dashes included in the plots denote the {\it rms} variations for
the cluster ensemble at each radius.
\label{fig:icl_rad}}
\end{figure}

Within the clusters, the radial dependence of the BCG+ICL luminosity fraction
is shown in Figure \ref{fig:icl_rad}, where we plot the BCG+ICL luminosity
fraction enclosed within different radii.  Combining the full ensemble of
clusters, the BCG+ICL luminosity fraction decreases smoothly from 65\% within
0.1\rtwo\ to 33\% within \rtwo.  This radial decline, which indicates that the
ICL is more concentrated than the galaxies, is also seen by
\citet{zibetti2005} and is predicted by the simulations of
\citet{murante2004}.

Lastly, we use a threshold of $\sigma=750$ \kms~--- roughly the median
velocity dispersion of an Abell richness class $R \ge 1$ cluster
\citep{zabludoff1990} --- to divide the data set into two subsamples (Figure
\ref{fig:icl}). The clusters with higher velocity dispersions exhibit lower
BCG+ICL luminosity fractions at all radii. To confirm that this result is not
driven purely by the BCG, we also compute the corresponding luminosity
fractions within circular annuli from 0 to $0.5$\rtwo. We find that at all
radii the fraction of the cluster luminosity contained in the ICL is lower for
the more massive clusters. Moreover, the rms scatter about the mean for the
BCG+ICL luminosity fractions is only 10\% for the high-$\sigma$ sample
compared with 20\% for the low-$\sigma$ sample within 0.5\rtwo.  Thus, for
clusters and groups with a dominant BCG, the BCG+ICL luminosity fractions are
smaller and more uniform in the more massive systems.

\subsection{Origin of the Baryon Trends}

We have identified two independent trends with mass that shed light on the
environmental dependencies of physical processes that distribute baryons among
the various phases. First, we find that the fraction of stars in the ICL grows
(and that in galaxies diminishes) with decreasing system mass.  This result is
predicated on the assumption that we are not strongly biased in the type of
system observed, as discussed above. Even so, we now know that in at least
some low mass systems the ICL forms efficiently.  As a result, we can rule out
cluster-only processes, such as harassment \citep{moore1996,moore1998}, as the
dominant ICL formation mechanism, and find support for mechanisms that would
be more effective in lower mass systems, such as galaxy-galaxy interactions
\citep{merritt1984}.  Such tidal interactions, which could strip stars from
galaxies and build up the ICL, are favored in lower mass groups, because the
internal velocity dispersions of group galaxies are similar to the velocity
dispersion of the group as a whole.  If all low mass systems do have high ICL
luminosity fractions, and some are the building-blocks of clusters, then a
diluting mechanism, such as the infall of individual galaxies into clusters,
would be necessary to reduce the final ICL fraction of more massive systems.
Of course, it may be that not all low mass groups have high ICL fractions like
those in our sample. The low ICL fractions of high mass clusters may therefore
simply reflect their growth by the accretion of groups with lower ICL
fractions than we observe.  A related possibility is that the clusters accrete
and mix progenitor groups before they have time to form substantial ICL via
galaxy-galaxy interactions, whereas the ICL continues to grow in more isolated
groups.  Simulations of the formation of the ICL must reproduce these trends
and will lead to a better understanding of the details of the formation
mechanisms \citep{Willman2004,murante2004,SL2005,SL2006,Rudick2006}.

Second, we find that the baryon mass fraction across the range of systems for
which we have both optical and X-ray data is roughly constant over the mass
range $6\times10^{13}\ \mathrm{M}_\odot <$ \mfive$\ < 1\times10^{15}\
\mathrm{M}_\odot$.  This result requires a fairly well-tuned balance between
the X-ray gas mass fraction, which declines by a factor of two between systems
with \mfive\ of $10^{15} M_\odot$ and $10^{14} M_\odot$, and the stellar mass
fraction, which increases by a factor of four to compensate. The matching of
the two trends supports the suggestion \citep{bryan,lin2003} that the X-ray
gas mass fraction behavior is due to the decreased efficiency of turning gas
into stars in the more massive environments.  The reduction in star formation
efficicency in higher mass systems may arise from the decreased efficiency of
tidal interactions among galaxies, the removal of the gaseous reservoirs of
galaxies by interactions with the intracluster medium, or increased feedback
processess that quench star formation.  As usual, interpreting such trends is
complicated by the fact that today's lower mass systems are not necessarily
similar to the lower mass antecedents of today's higher mass systems.

\section{Conclusions}
\label{sec:conclusions}

We have conducted a study of 23 nearby galaxy clusters and groups with a
dominant, early-type brightest cluster galaxy.  For this sample we determine
the luminosities of the different stellar constituents (including stars
associated with cluster galaxies and the intracluster stars) as a function of
cluster velocity dispersion ($\sigma$), cluster mass (\mfive), and
cluster-centric radius.  Combining these results with current measurements of
the X-ray gas mass as a function of $\sigma$, we calculate the dependence of
the baryon budget on environment. We reach the following conclusions:

1.  The observed total baryon mass fraction within \rfive\ is constant
for systems with \mfive\ between $6\times10^{13}$ and $10^{15}$ M$_\odot$.
Although the mean baryon mass fraction is formally $>3\sigma$ lower than the
WMAP measurement, both the observational systematic uncertainties --- which
dominate the formal random errors --- and the predicted deficit of baryons in
clusters relative to the universal value are on the order of the observed
baryon shortfall. Thus, we conclude that there is no compelling evidence for
undetected baryons at this time.

2. The inclusion of the combined light from the brightest cluster galaxy and
the intracluster stars, which we refer to as the BCG+ICL component, becomes
increasingly important in the baryon budget as the system mass decreases. For
systems with \mfive $= 10^{14} M_\odot$, the BCG+ICL component is nearly 35\%
of the total stellar luminosity within \rfive, and the mass of the stellar
component including the BCG+ICL and the cluster galaxies is comparable to that
of the X-ray gas component.  Consequently, the increase in X-ray gas mass
fraction with increasing \mfive\ is directly reflected by a decrease in the
total stellar mass fraction.  The matching of the falling stellar mass and
rising gas mass with total mass supports the suggestion made previously
\citep{bryan, lin2003} that the star formation efficiency decreases with
increasing system mass.

3. Considering only the stellar components, we observe that the fraction of
the total stellar luminosity in the BCG+ICL decreases with system velocity
dispersion. The origin of this trend is that the luminosity of the BCG+ICL
component changes more slowly with cluster mass than does the luminosity of
cluster galaxies, increasing by only a factor of two between 400 \kms~ and
1000 \kms. The presence of large BCG+ICL luminosity fractions in the lower
mass systems demonstrates that the rich cluster environment is not required
for production of ``intracluster'' stars.

The decrease in the importance of the BCG+ICL component, and in particular the
dominant ICL component, with increasing cluster mass is consistent with more
efficient stripping of stars from galaxies -- presumably via galaxy-galaxy
interactions -- in the lower mass systems.  Although it remains unclear the
degree to which our sample selection, which favored systems with central,
giant elliptical galaxies, impacts this result, it is noteworthy that
\citet{purcell2007} predict precisely the type of relation that we observe
using a simple analytic model for BCG+ICL formation from satellite accretion
and disruption. A similar model is also able to reproduce the relative
luminosities that we observe for the BCG and ICL \citep{conroy2007}, implying
that this basic physical picture can largely explain two significant results
from our work without invoking significant bias due to sample selection.

4.  The mean BCG+ICL luminosity fraction for our sample decreases
monotonically from 65\% within 0.1\rtwo\ to 33\% within \rtwo. This trend is
qualitatively similar to that seen in the stacked sample of
\citet{zibetti2005}.

5. Interior to \rtwo, we find that the BCG+ICL optical luminosity fraction is
inversely correlated with cluster velocity dispersion. Our mean value of 33\%
is somewhat larger than one might expect based upon the work of
\citet{zibetti2005}, who found a similar luminosity fraction of 32\% within a
smaller, 500 kpc radius. This difference likely reflects a difference between
the systems included in the two studies.  For clusters with $\sigma>750$
\kms~, we obtain a mean value of 23\% within \rtwo, which is consistent with
both the \citet{zibetti2005} results and simulated BCG+ICL luminosity
fractions within \rtwo\ for rich clusters \citep{Willman2004}.

\acknowledgements 
The authors thank the referee, Stefano Ettori, for excellent suggestions that
significantly improved the paper.  AHG also thanks Andrey Kravtsov and
Yen-Ting Lin for insightful discussions regarding this project.  AIZ is
supported by NSF grant AST-0206084 and NASA LTSA grant NAG5-11108.  DZ
acknowledges financial support for this work from a Guggenheim fellowship and
NASA LTSA grant 04-0000-0041. AIZ and DZ thank the KITP for its hospitality
and financial support through the National Science Foundation grant
PHY99-07949, and generous support from the NYU Physics department and Center
for Cosmology and Particle Physics during their sabbatical there.

\bibliographystyle{apj}
\bibliography{ms}

\end{document}